\documentclass[twocolumn,prd,aps,showpacs]{revtex4}

\newcommand{\be}{\begin{equation}}
\newcommand{\ee}{\end{equation}}
\newcommand{\rf}[1]{(\ref{eq:#1})}
\begin{document}

\title{Covariant Analysis of the Experimental Constraints on the Brane-world}

\author{M. D. Maia}
\affiliation{Universidade de Bras\'{\i}lia, Instituto de F\'{\i}sica, 70919-970,
Bras\'{\i}lia, DF,  maia@fis.unb.br}
\author{E. M. Monte}\email{edmundo@fisica.ufpb.br}
\affiliation{Departamento de F\'\i sica, Universidade Federal da Para\'\i ba,
58059-970, Jo\~ao Pessoa, PB, Brasil}

\begin{abstract}
Some    observational constraints
on the brane-world  based on  predictions 
from specific models in   five dimensions, have been  recently reported,  both on local  and  cosmological  scales. In order to  
identify the origins   of these constraints,  the equations 
of motion of the brane-world  are translated to the most general,  model-independent (or ``covariant"),  formulation of the theory, based only on the  Einstein-Hilbert action for the bulk geometry, the confinement of the  standard gauge interactions and the exclusive probing of the extra dimensions by the gravitational field. In the case of the  binary  pulsar PSR1913+16, it is  found that gravi-vectors and  gravi-scalars  do not appear in the covariant equations, but they are  replaced by   vector and  scalar fields related to  the extrinsic  curvature of the brane-world.
Only the latter  one  impose a  condition on the  binary pulsar 
orbits.   A general solution for this problem is proposed, based on results from differential geometry, suggesting  a stable bulk geometry, whose existence requires  higher dimensions.    On the cosmological  scale, it is  shown that the  high energy inflation  constraint originating from   the  square of the energy density term in the modified  Friedman's equation is  mainly due to the 
assumption of the reflection symmetry across the brane-world. 
It is shown that this symmetry is not consistent with the regularity of the brane-world.  These results suggest  that   the   two constraints can be lifted  by increasing the number of  extra dimensions.
 \end{abstract}


\maketitle

\section{The Constraints }

When a  theory  is
faced  with observational   constraints, it becomes necessary  to verify  the extent  in which  its  predictions depend on the properties of used models.  Only when those predictions are covariant in the sense of  being model independent  we may  convince ourselves  that the  constraints    pose a real
problem  to the theory in itself.  This appears to be  the  current situation in  brane-world   theory in five-dimensions,  where  several  models  have  been recently checked against precision
astrophysical and cosmological observational  constraints.  

 In particular, the  interference of linear gravitational waves  generated by the bulk geometry  over the  quadrupole  formula for  the binary pulsar PSR1913+16, predicts an  error of about $20\%$,  against an 
 observed  error  of just  $0.5\%$  \cite{Durrer}.
In the   cosmological scale, different models lead to  
a  modified  Friedman's  equation depending on the square of the energy density, whose effect is to produce  a  slowdown of the high energy  inflation, in  disagreement  with the    recent  data from the WMAP/SDSS/2dF surveys \cite{Liddle,Bohem}.

 The purpose of this paper is find  why these constraints occur,  by  rewriting them in the 
  most general  model-independent  formulation 
   (sometimes referred to as  ``covariant" formulation) 
 and, whenever possible  propose  solutions.  The   model-independent  formulation of the brane-world is   characterized  only by the Einstein-Hilbert action  for the bulk geometry,  the  confinement of  the standard  gauge interactions and   the  exclusive probing  of  the extra dimensions by the gravitational field  at     the TeV scale  of  energy \cite{ADD}.  The  result is  a  set of bare equations of motion,   to  which we may add specific   model properties  afterwards (These are in fact  well known  equations which have  been  extensively  applied  to the 
 construction of diverse   models in   five dimensions \cite{Maeda,Roy,Reza,Sahni,Tsujikawa,Gong,DGP}). 

We find  that   gravi-vectors and  gravi-scalars   do not appear    in the covariant equations, but  they are  replaced  by the  components of the  extrinsic curvature and  its  contraction, 
satisfying  equivalent  vectors  and  scalar  equations.
Therefore, using  the same  conditions  as in  \cite{Durrer}, we obtain  essentially  the same  graviton equations and   a  scalar  condition  which   interferes with the   binary pulsar  orbits.

A general  solution  for  such type of  constraint  is proposed,    by   increasing the number of  extra dimensions  in  the covariant formulation,   until  the bulk  
becomes    stable in the sense that its metric is   not perturbable,  regardless  of what is  happening  with  the  brane-world embedded  on it.

 On the  other hand,   we have  found that the primary source of  the  high energy inflation constraint is  in most cases the   $Z_2$  symmetry  across  the brane-world.  We  show  that  although  this  result can be implemented in  the covariant formulation on a specific brane-world, in  general the $Z_2$  symmetry  is  not  consistent with the   regularity theorems  required by its perturbations.

 \section{The Covariant equations of  motion  in Arbitrary Dimensions}

The four-dimensional brane-world  can be seen  as  the result  of  the motion of  a  3-brane  embedded in a $D$-dimensional  bulk,  $D=4 +N$,   whose geometry  is  defined by the   Einstein-Hilbert action integral$^{[1]}$\footnotetext[1]{Notation: Curly curvature components ${\mathcal{R}}_{...}$ 
refer to the bulk, while straight curvature components ${R}_{...}$ refer to the brane-world. Capital Latin indices run from 1 to D. Small case
Latin indices refer to the extra dimensions only, running from 5 to D. All Greek indices refer to the 
brane-world, counting from 1 to 4. An overbar denotes an
object of  a fixed background brane-world geometry.  The semicolon denotes the covariant derivative with respect to the brane-world metric  $g_{\mu\nu}$. } 
\begin{equation}
\int {\mathcal{R}}\sqrt{-{\cal G}}d^{D}v=\alpha_{*}
\int {\cal L}^{*}\sqrt{-{\cal G}}d^{D}v
\label{eq:bulkEH}
\end{equation}
 where $\alpha_{*}$ is the bulk energy scale  and ${\cal L}^{*}$ is a 
source Lagrangian  usually describing   gauge fields and ordinary matter confined  to the brane-world.
Taking the variation of this  action  with respect to the bulk metric ${\cal G}_{AB}$ we obtain the bulk Einstein's equations
 \begin{equation}
 {\cal R}_{AB} -\frac{1}{2}{\cal R}{\cal G}_{AB}=\alpha_{*}T^{*}_{AB} \label{eq:bulkEE}
 \end{equation}

The confinement hypothesis  states that 
the standard   gauge fields and  ordinary  matter remain trapped in a 3-brane, or better, in the 4-dimensional manifold spanned by its motion in the  bulk. On the other hand, 
the exclusive probing  of the extra dimensions by  TeV gravitons  say that  the geometry of that  manifold
present oscillation modes at that energy scale. Thus,  these two postulates require  a  submanifold structure, the brane-world, which   remains  always  embedded in the bulk.

The  embedding
of a manifold into another can be realized in many different ways and the choice of one or another
depends on what it  is  supposed to do. The action principle \rf{bulkEH}  suggests  that  the four-dimensional gravitational field  is  induced by  that of the bulk,  and the  simplest realization of such   induction  is through a local and  isometric embedding.

 A very common simplification  consists in assuming 
 that the  embedding functions are analytic in the sense that they are representable by convergent positive power series. This  type of  embedding is   is useful to prove theorems in mathematics,  but it bypasses some of the 
  differentiable properties required by the   dynamics
  of the brane-world. Thus, except in some particular  instances,  the  analytic embedding  is  not suitable for the  brane-world whose  geometry represents  a   high  energy  field  (at the least at the   TeV scale).
 The more general differentiable embedding,  obtained via  differentiable perturbations of  a  given background, is briefly reviewed below, mostly extracted  from the classic literature on this subject  \cite{Eisenhart,Nash,Greene}.

The differentiable embedding of a given manifold $\bar{V}_{4}$ with metric $\bar{g}_{\mu\nu}$ in an arbitrary
 bulk $V_{D}$, $D=4+N$,  with metric ${\mathcal{G}}_{AB}$, is given by $D$ differentiable maps
$\bar{\mathcal{X}}^{A}:\bar{V}_{4}\rightarrow V_{D}$,  satisfying  the (isometric)  embedding
equations
\begin{equation}
\bar{\mathcal{X}}_{,\alpha}^{A}\bar{\mathcal{X}}_{,\beta}^{B}
{\mathcal{G}}_{AB}\!=\!\bar{g}_{\alpha\beta},\;\bar{\mathcal{X}}_{,\alpha}^{A}\bar{\eta}_{a}^{B}
{\mathcal{G}}_{AB}\!=\!0,\;\bar{\eta}_{a}^{A}\bar{\eta}_{b}^{B}{\mathcal{G}}_{AB}\!=\!\epsilon_{a}
\delta_{ab}\!\!\label{eq:X}
\end{equation}
 where $\bar{\eta}_{a}$ denotes the components of the
 $N=D-4$  orthogonal 
vectors  normal to $\bar{V}_{4}$, and $\epsilon_{a}=\pm1$ correspond to the  two possible  signatures of each extra
dimension. Once we have the embedding of that  particular  $\bar{V}_4$, we may deform  (or perturb)
it  along an arbitrary direction $\zeta$ in the bulk, given by  the Lie derivative of the embedding
coordinates
\begin{equation}
{\mathcal{Z}}^{A}=\bar{X}^{A}+(\pounds_{\zeta}\bar{\mathcal{{X}}})^{A}  \label{eq:perturbation}
\end{equation}
To avoid possible  coordinate gauges which could  trigger  false perturbations, as in  \cite{Nash} we consider the
deformations along the unit normals $\eta_{a}$,  parameterized by the extra coordinates $y^{a}$. In
this case, the components of the deformed embedding  functions 
${\mathcal{Z}}^{A}=\bar{X}^{A}+y^{a}\eta_{a}^{A}$ must satisfy embedding equations similar to \rf{X}, with the difference that now they depend  on  $y^{a}$. Using  \rf{X} and  \rf{perturbation}, we obtain  the  geometry  of the perturbed  manifold
\begin{eqnarray}
g_{\mu\nu}(x,y) & = &
{\mathcal{Z}}_{,\mu}^{A}{\mathcal{Z}}_{,\nu}^{B}{\mathcal{G}}_{\!
AB}=\bar{g}_{\mu\nu}\!\!- 2y^{a}\bar{k}_{\mu\nu a}  + \nonumber \\
 &&
y^{a}y^{b}[\bar{g}^{\alpha\beta}\bar{k}_{\mu\alpha a}\bar{k}_{\nu\beta b}+g^{cd}\bar{A}_{\mu
ca}\bar{A}_{\nu db}],\label{eq:gmunu}\\
g_{\mu a}(x,y) & = & 
{\mathcal{Z}}_{,\mu}^{A}{\eta}_{a}^{B}{\mathcal{G}}_{\! AB}=\!\! y^{a}A_{\mu ab},  \label{eq:gmua}\\
g_{ab}(x,y) & = & 
{\eta}_{a}^{A}{\eta}_{b}^{B}{\mathcal{G}}_{\! AB}=\!\!\epsilon_{a}\delta_{\mu 
a},\epsilon_{a}\!\!=\!\!\pm1,g^{ab}g_{bc}\!=\!\delta_{c}^{a}\label{eq:gab}
\end{eqnarray}
where  the  first equals  signs  show  that  these  are  the components of the  bulk metric ${\cal G}_{AB}$  evaluated in the  embedding vielbein  $\{{\cal Z}^{A}_{,\mu}, \eta^{A}_{a} \}$

 In addition to the  metric  components  we  have also
 the  extrinsic curvature and the  ''torsion" vector, respectively given by
\begin{eqnarray}k_{\mu\nu a}(x,y) & = &
-\eta_{a,\mu}^{A}{\mathcal{Z}}_{,\nu}^{B}{\mathcal{G}}_{AB}\label{eq:k}=\\ 
&&\phantom{x}\hspace{-10mm}\bar{k}_{\mu\nu
a}\!\!-y^{b}\bar{g}^{\alpha\beta}\bar{k}_{\mu\alpha a}\bar{k}_{\nu\beta
b}-\!\!{g}^{cd}y^{b}\bar{A}_{\mu ca}\bar{A}_{\nu db},\hspace{3mm}\nonumber \\
A_{\mu ab}(x,y) & = &
\eta_{a,\mu}^{A}\eta_{b}^{B}{\mathcal{G}}_{AB}\!\!=\!\!\bar{A}_{\mu ab}(x)\label{eq:A}
\end{eqnarray}
Notice that  \rf{gmunu} and \rf{k}  imply that
the extrinsic  curvature also 
propagates in the bulk, according to 
 York's relation (extended to the extra variables
$y^{a}$):
\begin{equation}
k_{\mu\nu a}=-\frac{1}{2}\frac{\partial g_{\mu\nu}}{\partial y^{a}} \label{eq:york}
\end{equation}
However, from  \rf{A} it follows  that   $A_{\mu ab}$ does not propagate at all in the bulk.
Finally, defining
$h^{AB}=g^{\mu\nu}Z_{,\mu}^{A}Z_{,\nu}^{B}$, with
inverse $h_{AB}={\mathcal{G}}_{AM}{\mathcal{G}}_{BN}h^{MN}$,  we obtain
from \rf{gmunu}-\rf{gab}
\begin{equation}
h^{AB}={\mathcal{G}}^{AB}-g^{ab}\eta_{a}^{A}\eta_{b}^{B}\label{eq:invert}
\end{equation}

In order to guarantee that  the perturbed manifold
remains embedded in the same  bulk, 
the   Riemann  tensor ${\mathcal{R}}_{ABCD}$  must be independent of the  hypersurface on which its components are  expressed. Therefore, using the vielbein $\{{\mathcal{Z}}_{,\mu}^{A},\eta_{a}^{A}\}$ defined by the  perturbed hypersurface, the  components 
 of  that tensor give the required  condition.  After
eliminating the  redundant expressions, the remaining equations are the well known  Gauss,  Codazzi
and Ricci equations: 
\begin{eqnarray}
{\mathcal{R}}_{ABCD}{\mathcal{Z}}_{,\mu}^{A}{\mathcal{Z}}_{,\nu}^{B}
{\mathcal{Z}}_{,\rho}^{C}{\mathcal{Z}}_{,\sigma}^{D}  = &&R_{\mu\nu\rho\sigma}\!\!-\!\!2g^{cd}k_{\mu[\rho c}k_{\sigma]\nu d}\label{eq:Gauss}\\
{\mathcal{R}}_{ABCD}{\mathcal{Z}}_{,\mu}^{A}\eta_{a}^{B}{\mathcal{Z}}_{,\nu}^{C}{
\mathcal{Z}}_{,\rho}^{D}  = && 2k_{\mu[\nu a;\rho]}\!\!-\!\!2g^{cd}A_{[\rho 
ca}k_{\mu\nu]d}\label{eq:Codazzi}\\
{\mathcal{R}}_{ABCD}\eta_{a}^{A}\eta_{b}^{B}{\mathcal{Z}}_{,\mu}^{
C}{\mathcal{Z}}_{,\nu}^{D} =  && -2A_{[\mu ab;\nu]}\!\!-\!\!2g^{cd}A_{[\mu ca}A_{\nu]db}\nonumber\\ 
 -&&2g^{\alpha\beta}k_{[\mu\alpha a}k_{\nu]\beta b}\label{eq:Ricci}
\end{eqnarray}
where  brackets  apply  to the adjoining  indices only.

It is  clear that the  equations of  motion  of the brane-world must also be compatible with those equations. In fact they can be derived  from   the contraction of  \rf{Gauss} with
$g^{\mu\nu}$.  After using\rf{invert} we
obtain
\begin{eqnarray}
{\mathcal{R}}_{AB}Z_{,\mu}^{A}Z_{,\nu}^{B}=R_{\mu\nu}\!\!- g^{cd}(g^{\alpha\beta}k_{\mu\alpha c}k_{\nu\beta d}\!\! & - &\!\!
H_{c}k_{\mu\nu d})-\nonumber \\ 
&&\!\! 
\phantom{x}\hspace{-3cm}g^{ab}{\mathcal{R}}_{ABCD}\eta_{a}^{A}Z_{,\mu}^{B}Z_{,\nu}^{C}\eta_{b}^{D}
\label{eq:RicciT}
\end{eqnarray}
where we have denoted $H_{a}=g^{\mu\nu}k_{\mu\nu a}$.
A  further contraction with $g^{\mu\nu}$ gives the Ricci
scalar
\begin{eqnarray}
{\mathcal{R}}= R -(K^{2}-H^{2}) & + &
2g^{ab}{\mathcal{R}}_{AB}\eta_{a}^{A}\eta_{b}^{B}\nonumber -\\ 
&&\phantom{x}\hspace{-2cm}
g^{ad}g^{bc}{\mathcal{R}}_{ABCD}\eta_{a}^{A}\eta_{b}^{B}\eta_{c}^{C}\eta_{d}^{D}\label{eq:RICCI}
\end{eqnarray}
where
 $K^{2}=g^{ab}k^{\mu\nu}{}_{a}k_{\mu\nu b}$ and
 $H^{2}=g^{ab}H_{a}H_{b}$.  
From  \rf{RICCI}   we  may write 
 the same Einstein-Hilbert action   \rf{bulkEH}, but now expressed in terms 
 of the intrinsic and extrinsic 
brane-world geometry
\begin{eqnarray}
&& \phantom{x}\hspace{-10mm} \int{\cal R}\sqrt{-{\cal G}}d^{D}\!\!v=
\int[R-K^{2}+H^{2}]\sqrt{-{\cal G}}d^{D}\!\!v+\nonumber\\ 
 &&\phantom{x}\hspace{-12mm}\int \!\![2g^{ab}{\cal R}_{AB}\eta^{A}_{a}\eta^{B}_{b}\!\!-\!\!g^{ad}g^{bc}{\cal R}_{ABCD}\eta_{a}^{A}\eta_{b}^{B}\eta_{c}^{C}\eta_{d}^{D}]\sqrt{\!\!-{\cal G}}d^{D}v\nonumber\\
&& = \alpha_{*}\int {\cal L}^{*}d^{D}v \label{eq:EH}
\end{eqnarray}
If  wished,  at the  level of the  variational principle, we may add terms such as   boundaries and  cosmological  constant. 

 Taking the    variation   of  \rf{EH}  with respect  to
 the separate metric components  $ g_{\mu\nu}$,   $ g_{\mu  a}$  and  to $ g_{ab}$,  and
 denoting by $ T_{\mu\nu}^{*}=T^{*}_{AB}{\cal Z}^{A}_{,\mu}{\cal Z}^{B}_{,\nu}$, $ T_{\mu  a}^{*}=T^{*}_{AB}{\cal Z}^{A}_{,\mu}\eta^{B}_{a}$ and $T_{ab}^{*}=T^{*}_{AB}\eta^{A}_{a}\eta^{B}_{b}$   the corresponding  energy-momentum tensor  components derived from ${\cal L}^{*}$, 
 we obtain the  covariant equations  of  motion  consistent with 
  \rf{RicciT} and  \rf{RICCI}   
\begin{eqnarray}
&&\phantom{x}\hspace{-10mm}(R_{\mu\nu}\!\!-\!\!\frac{1}{2}R g_{\mu\nu})-Q_{\mu\nu}-
g^{ab}{\mathcal{R}}_{AB}\eta_{a}^{A}\eta_{b}^{B}g_{\mu\nu}\nonumber + \\
&&\phantom{x}\hspace{1cm}(W_{\mu\nu}-\frac{1}{2}W g_{\mu\nu})=\alpha_{*}T_{\mu\nu}^{*},  \label{eq:BE1}\\
 &&\phantom{x}\hspace{-10mm}k_{\mu a;\rho}^{\rho}\!\! -\!\!H_{a,\mu}  \!\!-\!\!\frac{1}{2}(R\!\!-\!\!K^{2}\!\!+\!\!H^{2})g_{a\mu}  \!\!+\!\! A_{\rho c a}k^{\rho c}_{\mu}\!\! -\!\!A_{\mu c  a}H^{c}-\nonumber\\
 &&
\phantom{x}\hspace{10mm} W_{\mu a}\!\! +\!\! g^{mn}{\cal R}_{AB}\eta^{A}_{m}\eta^{B}_{n}g_{\mu a}  =\alpha_{*} T_{\mu  a}^{*}\label{eq:BE2}\\
&&\phantom{x}\hspace{-10mm}R-K^{2} +H^{2} +\frac{N-1}{2}{\cal R} +\frac{1}{2}W =\alpha_{*}T^{*}
\label{eq:BE3}
\end{eqnarray}
where we have denoted
\begin{eqnarray}
Q_{\mu\nu}\!\! & = & \!\! g^{ab}k^{\rho}{}_{\mu a}k_{\rho\nu b}-H^{a}k_{\mu\nu
a}\!\!-\!\!\frac{1}{2}(K^{2}-H^{2})g_{\mu\nu}\label{eq:Qij}\\
W_{\mu\nu}\!\! & = & \!\!
g^{ad}{\mathcal{R}}_{ABCD}\eta_{a}^{A}{\mathcal{Z}}_{,\mu}^{B}{\mathcal{Z}}_{,\nu}^{C}\eta_{d}^{D}\\
W_{\mu  a}&=& g^{mn} {\cal R}_{ABCD}\eta^{A}_{a}\eta^{B}_{m}{\cal Z}^{A}_{,\mu}\eta^{D}_{n}\\ 
W & = & g^{ad}g^{bc}{\mathcal{R}}_{ABCD}\eta_{a}^{A}\eta_{b}^{B}\eta_{c}^{C}\eta_{d}^{D} 
\end{eqnarray}
Notice that when all  extrinsic  properties are  removed  in  \rf{BE1}  we recover the  usual  Einstein's equations in  four  dimensions with the appropriate  value of  $\alpha_{*}$. However, with the embedding 
the presence of the extrinsic curvature  components is unavoidable. In particular,  the tensor $Q_{\mu\nu}$ is quadratic in the extrinsic curvature and it is conserved in the sense that $Q^{\mu\nu}{}_{;\nu}=0$, as it can be directly verified \cite{MEMA}.

Notice also that the  above equations   are equivalent  to  \rf{bulkEE}.  The  difference  is  that 
the solutions of \rf{BE1}-\rf{BE3} represent  a perturbation generated family of
brane-worlds,  whose embedding  in the bulk is   already built in 
these  equations, while  in the case of  \rf{bulkEE},  the  embeddings of the  perturbations  need  to be  checked afterwards. This  difference  is   due to the  distinct choice of  dynamical variables  for the same  action.  In \rf{bulkEE}  the dynamical variables are the components of the bulk metric ${\cal G}_{AB}$ chosen  after a  metric ansatz,  while in  \rf{BE1}-\rf{BE3}  the  components of the same  metric \rf{gmunu}-\rf{gab}, are obtained in the embedding frame. This  difference may  indicate that the gravi-scalar and  gravi-vectors depend on  the  basic  brane-world  in which they  are  written.
Under  a  perturbation of that  brane-world these quantities may change.  In this respect it is enlightening to compare with another choice of  dynamical variables (${\cal Z}^{A}$) for the same  action,  made 
in the past,  producing a  weaker set  of  equations as  compared with \rf{bulkEE} \cite{RT}.

\section{Constraining the Five-dimensional Bulk}

Since all reported constraints refer to five 
dimensional models, in  this  section we restrict the previous analysis  to $D=5$, noting that in 
this case $A_{\mu\nu a}=0$ and $W=0$.
For notational simplicity  denote  $k_{\mu\nu5}=k_{\mu\nu}$. Also,  to  simplify  our  arguments,  we  fix  the bulk signature  
to be $(4,1)$  (That is $g^{55}=1$). Then, the  covariant equations  \rf{BE1}-\rf{BE3} simplify to 
\begin{eqnarray}  
&&\phantom{x}\hspace{-10mm}(R_{\mu\nu}\!\!-\!\!\frac{1}{2}Rg_{\mu\nu})\!\!-\!\!Q_{\mu\nu}\!\!-\!\!  
R_{AB}\eta^{A}\eta^{B}g_{\mu\nu}\!\!+ \!\!W_{\mu\nu}\!\!=\!\!\alpha_{*} T^{*}_{\mu\nu},
 \label{eq:BE51}\\ 
&& \phantom{x}\hspace{+8mm}k_{\mu ;\rho}^{\rho}  -H_{,\mu}=  \alpha_{*} T^{*}_{\mu 5},  \label{eq:BE52}\\
 &&\phantom{x}\hspace{+3mm}R-K^{2} +H^{2}  =-2\alpha_{*} T^{*}_{55}\label{eq:BE53}
\end{eqnarray}
In the following we apply these equations to the
analysis of the binary pulsar and the high energy inflation constraints. 
 \vspace{3mm}\\ 
1- \underline{Constraints  generated by bulk gravitational waves}\\ 
\vspace{3mm}

 The model used  in  \cite{Durrer} is  based on  a fairly general metric  ansatz  defined in  a cylindrical  bulk with an oscillating  radius, but without implementing  the  $Z_{2}$  symmetry. 
 Considering  the linear  expansion of the bulk metric 
\begin{equation}
{\mathcal{G}}_{AB}=\eta_{AB}+\gamma_{AB}\label{eq:linear}
\end{equation}
 and  applying the corresponding de Donder Gauge,  we obtain the   wave equations $\fbox{}^{2}\Psi_{AB}=
 \alpha_{*}T^{*}_{AB}$. Using a    coordinate  system in which
 these  equations split  as
 \begin{eqnarray}
&&\fbox{}^{2}\Psi_{\mu\nu}=\alpha_{*}T_{\mu\nu}^{*}\;\;\;\mbox{gravitons} \label{eq:S1}\\
&&\fbox{}^{2}\Psi_{\mu  5}=\alpha_{*}T_{\mu 5}^{*}\;\;\;\mbox{gravi-vector} \label{eq:S2}\\
&&\fbox{}^{2}\Psi_{55}=\alpha_{*}T_{55}^{*}\;\;\;\mbox{gravi-scalar}  \label{eq:S3}
	\end{eqnarray}
	By following  a perturbative approach,
 it was found that   the gravi-vector does not produce any appreciable consequence on the  orbits  of the  binary pulsar  PSR1913+16, but the gravi-scalar induce a  slowdown of its  period   by $-2.87\times10^{-12}$, instead of the known experimental value $-(2.408\pm 0.001)\times10^{-12}$ \cite{Durrer}. 
Although  these results  are  claimed to be coordinate gauge independent, it  is not  clear  the  extent  in  which they depend  on the chosen  bulk.

To see  how the  same problem translates to the 
 covariant formulation, consider  the same linear perturbation of the bulk geometry.  It is  a  simple matter to see from  \rf{gmunu}  that this  perturbation is  transferred to the brane-world  metric as   
 \[
 g_{\mu\nu}(x,y)=\eta_{\mu\nu}  +  \gamma_{\mu\nu}, \;\;\; \gamma_{\mu\nu} ={\cal Z}^{A}_{,\mu}{\cal Z}^{B}_{,\nu}\gamma_{AB}
 \]
and   from  \rf{york} 
we  obtain    $k_{\mu\nu}= -\partial  \gamma_{\mu\nu}/2\partial y$.  Since  $Q_{\mu\nu}$  is  quadratic in  $k_{\mu\nu}$, it  follows that  such term is  negligible in  the presence of   linear  terms on  $\gamma_{\mu\nu}$. Consequently, in the same  cylinder bulk and using   $T_{\mu\nu} =\bar{T}_{\mu\nu}\delta{(y)},\;\;  T_{\mu 5}=T_{55}=0$,  equation  \rf{BE51}
becomes  $\fbox{}^{2}\Psi_{\mu\nu}=  8\pi G \bar{T}_{\mu\nu}\delta{(y)} $, which is  essentially  the same  graviton  equation \rf{S1}.

On the other hand, instead of the equations  \rf{S1},  \rf{S2}, we  have the equations  \rf{BE52} which corresponds  to the trace of  Codazzi's  equations and  this will be  an  identity  in  a  bulk with  cylindrical topology. Finally, in the same  approximation  we   obtain  that  $H^{2}=0$ and  $K^{2}=0$, so  that  equation \rf{BE53} in the same de  Donder  gauge is equivalent to   $g^{\mu\nu}\fbox{}^{2}\Psi_{\mu\nu}=0$, implying that   
$T_{\mu\nu}$  must be trace-free. 

Therefore, we  obtain from the covariant equations of  motion a  set of  equations  which is   equivalent  to  the  equations in  \cite{Durrer}, suggesting that  the bulk generated  gravitational waves  in general  imply in a constraint  to the binary pulsar,  but its  effectiveness  must be revalued.

It is  interesting  to note  that the  massive  modes 
were  not taken  into consideration in  \cite{Durrer}. 
However,  when the   extra  dimension is  compact,    the tangent components of the wave functions can be  always harmonically expanded as
 \[
\Psi_{\mu\nu}=\sum_{n}\beta_{\mu\nu}^{(n)}\; 
e^{\frac{in\pi y}{\ell}}
\]
Thus, denoting $M_{n}=n^{2}\pi^{2}/\ell^{2}$ and the n-mode by $\Psi_{\mu\nu}^{(n)}$, the  wave equation can be written as  
\[
\fbox{}^{2} \Psi_{\mu\nu} =\sum_{n}(\eta^{\alpha\beta}\partial_{\alpha}\partial_{\beta}-M_{n}^{2})
\Psi_{\mu\nu}^{(n)}= 8\pi G T_{\mu\nu}
\]
Unless we   restrict ourselves to  the zero-mode, a  four-dimensional observer naturally interprets $M_{n}$ 
as a mass attached to the n-mode $\Psi_{\mu\nu}^{(n)}$. As it  has been  suggested  in  another contexts (as for example affecting the  bending of light rays from distant sources by the Sun \cite{Damour}),  depending on  the size of $\ell$,  such masses may  contribute to local gravity  effects on the brane-world,  including the  binary pulsar.   
\vspace{1mm}

The existence of the binary  pulsar constraint   suggests that  
the  bulk geometry  should be   algebraically  special, of  a non-radiative type. However,   it is   hard to explain such hypothesis on theoretical grounds. More realistically, the 
binary pulsar   constraint can  be  seen as  an  indication that  the dimension  of the bulk cannot be fixed to  five.
 Indeed,  as  a   manifold  the  bulk can also be  embedded in a larger  bulk. Thus, supposing that  
 we have initially a  five-dimensional flat bulk whose geometry is  perturbed, producing a new non-flat five-dimensional bulk, also  containing the  original brane-world.  The perturbed non-flat bulk can  be  embedded in, say, 
a six-dimensional flat bulk. If for some physical action the geometry of this six-dimensional bulk is 
again  perturbed, then it can be re-embedded in an even higher-dimensional flat bulk. 
Such re-embedding process can be repeated over and over, until reaching a sufficiently 
higher-dimensional flat bulk whose geometry is flat and stable in the sense that it does not  change 
regardless of the dynamical state of the  embedded  brane-world. According with such view, the binary pulsar  
is  suggesting that  $D=5$ is  not sufficient.

This   flat-stable  bulk can  be considered as the analogous to    the ground state  of  Kaluza-Klein theory,   represented  by the D-dimensional  Minkowski space solution of  Einstein's  equations. However, this analogy is only partial  because   Kaluza-Klein theory is  based on the  product topology  $V_4 \times  B_N$ and not on a dynamical  submanifold structure like in the brane-world case. Nonetheless,   $V_4$  can be  seen   as   locally embedded  in the  product  space  obeying the same  Einstein-Hilbert dynamics, leading to some  variations of the Kaluza-Klein theory  which closely resembles the brane-world theory \cite{MaiaK,Visser,Gibbons,Kaku}.

In the general case, the  existence of such flat-stable 
 bulk was established originally by Nash and later on generalized to non positive metrics by Greene
\cite{Nash,Greene}, producing the  general  expression for the bulk dimension: $D=n(n+3)/2$. For a  3-brane  this gives  $D=9$ and   for  the  brane-world we obtain 
$D=14$.  It is  important  to realize  that  this  has  nothing to do  with  supergravity  which would impose  a  $D=11$  limit.  Yet, it has been  noted that a  $14$-dimensional  bulk with signature   $(13,1)$  can also embed some interesting super-algebras  \cite{Bars}.
Also  it should be remenbered  that the use of the  extra embedding dimensions as  a possible generator  of gauge symmetries has  been proposed 
long ago \cite{Neeman,Joseph}, suggesting   an  $SO(10)$ based  GUT model on the the brane-world.    

\vspace{1mm}
 2 - \underbar{High Energy Inflation Constraint}
\vspace{1mm}

Here  we have a  different problem  resulting from the 
presence of the  $\rho^{2}$  term   in the brane-world modified  Friedman's  equation.  Accordingly, 
 the presence of such  term implies  in a 
slowing down of the high energy inflation, inconsistently with the anisotropy data from the 
WMAP/SDSS/2dF experiments \cite{Liddle,Bohem}. 

In  order  to understand  how  the   $\rho^{2}$ term appears  in Friedman's equation consider again Einstein's equations for the bulk, now  written as 
\begin{equation}
{\mathcal{R}}_{AB}=\alpha_{*}(T_{AB}^{*}-\frac{1}{3}T^{*}g_{AB}) 
\label{eq:bulkEE2}
\end{equation}

Using a Gaussian  normal coordinates on the brane-world, in which the   metric components  are $g_{\mu\nu}$, $g_{\mu 5}=0$ and  $g_{55}=1$,  the tangent components ${\cal R}_{\mu\nu}$  of the above equations are
(after using \rf{york})
 \begin{eqnarray}
&&\phantom{x}\hspace{-10mm} R_{\mu\nu}  -\!\frac{\partial k_{\mu\nu}}{\partial y}\!-2k_{\mu}^{\rho}k_{\rho\nu}  +h k_{\mu\nu} =\alpha_{*}(T^{*}_{\mu\nu}-\frac{1}{3}T^{*}g_{\mu\nu})
 \label{eq:R+}
 \end{eqnarray}
Taking  the  brane-world as  
   a  boundary, separating two  regions  of the bulk, labeled   by   $+$ and $-$  respectively, the difference  between the  above components, calculated  on each side of the brane-world for  $y\rightarrow 0$ is  zero because  there is  no real    distinction of the Riemann geometry  of the bulk as  seen from each side.   This  situation  changes  when the   $Z_{2}$ symmetry  is assumed  across the
 brane-world $\bar{V}_{4}$, so that  the brane  acts as  a mirror. In this case,
   an object   that senses the  extra  dimension in one side is  mirrored by $\bar{V}_{4}$  to its   image.  This is  the case of  the extrinsic curvature  which measure the  
tangent component of  the  variation of the  normal vector $\eta$, when its  foot is displaced on the brane-world. From the mirror  image of  this variation  we obtain  $k^{+}_{\mu\nu}=-k^{-}_{\mu\nu}$.
 Using   the  mean value theorem for the derivative of  $k_{\mu\nu}$ with respect to  $y$,  we also find
 that  (with $k_{\mu\nu}^{+} =k_{\mu\nu}$)
 \[
 -(\frac{\partial 
k_{\mu\nu}}{\partial y})^{+} +(\frac{\partial 
k_{\mu\nu}}{\partial y})^{-}  =  -2\frac{k_{\mu\nu}}{y} 
\]
Denoting   $[X]=X^{+}-X^{-}$, and  $X=\bar{X}(x)\delta(y)$, under  the  $Z_2$ symmetry it follows that in the limit   $y \rightarrow 0$  
\[
|y|[X]  =
\int_{-y}^{y}|y|\bar{X}\delta'(y) dy  +\int_{-y}^{y}\frac{y}{|y|}\bar{X}\delta (y) dy  =2\bar{X}
\]
Replacing these expressions  in  the  difference of  the expression \rf{R+}  calculated in both sides, for  $X=\bar{T}_{\mu\nu}$,    we obtain at  $y=0$  the  Israel-Lanczos  condition
\begin{equation}
\bar{k}_{\mu\nu}=-\alpha_{*}(\bar{T}_{\mu\nu}-\frac{1}{3}\bar{T}\bar{g}_{\mu\nu})
\label{eq:israel}
\end{equation}
Therefore,  this condition follows from  Einstein's  bulk equations \rf{bulkEE}, plus the  $Z_2$ symmetry, plus the delta function confinement  of $ T^{*}_{\mu\nu}$.

The  $Z_2$  symmetry across  the brane-world was originally  motivated  by the Horava-Witten theory to  compactify  the  11-dimensional M-theory  to the  product  topology  $V_{10}\times S^1 /Z_2$.   However,  when the same principle   is  transposed  to the brane-world   theory  based on the Einstein-Hilbert  action  in  five dimensions,  the  bulk  becomes  orbifold  compactified  to $V_4 \times S^{1}/Z_2$ with the identification $-y\rightarrow y$ \cite{RS,Lupercio,Kawamura,Anne}. However, $S^{1}/Z_2$  is not  a manifold  and  some of  the conditions required for the  differentiable embedding   fail  to apply. Actually, under the  $Z_2$ symmetry,  all
 perturbations of the brane-world have  a mirror perturbation on the  opposite  side of the background,
 with the  derivatives of the normal having  opposite signs. Consequently, the   regularity of the embedding functions is  not generally  defined and the mentioned theorems of Nash and Greene, which depend on  extended regularity,  fail to apply. In other words,  
 the  implementation of the $Z_2$  symmetry  is not  completely  consistent with   covariant formulation of
 the  brane-world. However,  if we restrict   this to  the background  brane-world  $\bar{V}_{4}$,  as it is the case of \rf{israel}, then  it does not  really  matter  because this  background  was assumed  to be   embedded in the first place.

  The implications  of this symmetry, or  of  \rf{israel}, to the 
    high energy inflation  is  as follows:
Taking $\bar{T}_{\mu\nu}$ as   the 
energy-momentum tensor of the confined perfect fluid
in  \rf{israel} and  replacing  $\bar{k}_{\mu\nu}$ in the expression of  $Q_{\mu\nu}$   in  \rf{BE51}, then 
Friedman's equation   becomes  modified  by the addition of  a  squared energy density $\rho^{2}$  term \cite{Maeda,MaiaFr}.   Therefore, the high energy constraint  remains  valid in the  background geometry
only  and  it  is suggesting that  the  $Z_2$ symmetry is not  consistent with the perturbations of that background along the extra dimensions.   It is possible to replace  the  $Z_2$  symmetry  by alternative 
conditions applied to the  geometry of the brane-world. In these   cases the  effects of the  alternative conditions on the  differentiable structure need to be examined  separately \cite{Anne,Carter}.

\begin{center}
\textbf{Concluding Remarks}
\end{center}

The binary pulsar   PSR1913+16  has proven to be  a  valuable tool to  test alternative gravitational theories \cite{Will}.  Therefore,  it  should be  applied as a  test  to  the brane-world proposal, as it was done in \cite{Durrer}.   However,  to be  certain,  the  evaluation  must be done in a  model-independent fashion, as long as  we   have a consensus on  what is meant by model-independent.  In the present note  we  have   derived  
the  equations of motion for  a general brane-world in an arbitrary dimensions,  based only on the  Einstein-Hilbert action principle for the bulk geometry,  the  confinement hypothesis and the exclusive probing of the extra dimensions  by the  four-dimensional gravitational field,  leaving aside any  specific model property.  The probing energy   can be  fixed at  the  TeV scale,  but here this was generally  assigned to a  constant  $\alpha_{*}$. 

Taking the  dynamical variables as  the bulk metric  components in the  embedding frame, we  found  the    
equations of  motion  \rf{BE1}-\rf{BE3} which describe an  embedded brane-world  in the bulk,   under the very reasonable conditions that their embedding functions  remain  differentiable and  regular. It follows from these equations  that 
 the  five-dimensional bulk  may not be sufficient to  embed all possible  configurations of  a dynamically evolving brane-world.
 Yet, models defined in the five-dimensional bulks  have  become so  popular in the past five years that it almost become a  synonymous of the  brane-world  theory. The binary  pulsar  constraint   corroborate  that  such limitation on the bulk dimension cannot  be maintained.
 
 This conclusion  is    supported by the high energy inflation constraint. Indeed, we  have shown  that  the $Z_2$  symmetry  adopted in the most of the five-dimensional brane-world cosmologies
 is the primary  source of the  cause of the  high energy inflation constraint.  As it happens,  this  symmetry implies that all perturbations  of  a brane-world has a  mirror image  with respect to  $y=0$ and this  implies that  the  tangent  components of the  normal vector derivatives  have opposite directions. 
 When this  is  used as  an  orbifold  compactification,  the regularity of the embedding functions  becomes   undefined and consequently  the  differentiable embeddings  are compromised.  The high energy inflation constraint is  telling us that the  $Z_2$  symmetry and  
 the  orbifold  compactification  must not be present on the brane-world.  
Therefore,   the  two  constraints  suggests by   different  ways  that   the brane-world program requires
more than  five dimensions.

\begin{center}Acknowledgments \end{center}

The author wishes to thank the very helpful, comments
and suggestions of R. Durrer and R. Tavakol on a previous version of this note, and
to M. Reboucas for a discussion on the higher-dimensional Petrov classification.


\begin{thebibliography}{10}

\bibitem{Durrer} R. Durrer \& P. Kocian, hep-th/0305181 
\bibitem{Liddle} S. Tsujikawa \& A. R. Liddle, astro-ph/0312162 
\bibitem{Bohem}  C. Ringeval, T. Boehm, R. Durrer hep-th/0307100 
\bibitem{ADD}   N. Arkani-Hamed et Al,  Phys. Lett. {\bf{B429}}, 263,  (1998), Phys. Rev. Lett. {84}, 586, (2000)
\bibitem{Maeda}    K. Maeda et al, Phys. Rev.  {\bf{D62}}, 024012 (2000)
\bibitem{Roy} R. Maartens, gr-qc/0312059 
\bibitem{Reza}E. Anderson \& R. Tavakol. Class.Quant.Grav. \textbf{20}, L267, (2003),
gr-qc/0305013 
\bibitem{Sahni}V. Sahni \& Y. Shtanov,  astro-ph/0202346 
\bibitem{Tsujikawa} S.  S. Tsujikawa et al, astro-ph/0406078
\bibitem{Gong} Y. Gong \& Chang-kui Duan,
astro-ph/0401530
\bibitem{DGP}G. Dvali, G. Gabedadze, M. Porrati, Phys. Lett. \underbar{B 485}, 208, (2000). hep-th/0002190, hep-th/0005016. 
\bibitem{Eisenhart}  L. P. Eisenhart, {\em Riemannian Geometry}, Princeton U.P.  (1966).
\bibitem{Nash} J. Nash, Ann. Maths.  \underbar{63}, 20, (1956) 
\bibitem{Greene}R. Greene, Mem, Amer. Math. Soc. \underbar{97}, (1970) 
\bibitem{MEMA}M.D. Maia et al, astro-ph/0403072

\bibitem{Damour} T. Damour et al, hep-th/0212155 
\bibitem{RT} T. Regge \&  C. Teitelboim,  Proc. II Marcel Grossman Meeting, ICTP  (1975),  S.Deser,  F.A.E. Pirani, D.C. Robinson, Phys. Rev. D14, 3301 (1976)
\bibitem{MaiaK}  M. D. Maia, Phys. Rev. D. {\bf{31}}, 262 (1985)
 \bibitem{Visser}  M. Visser,  Phys. Lett \underline{B159}, 22, (1985)
 \bibitem{Gibbons},  G. W.  Gibbons \&  D. L.Wiltshire, Nucl. Phys.  \underline{B287}, 717, (1987)
 \bibitem{Kaku},  M.Kaku \& J. Lykken, Nucl. Phys \underline{B268},  489  (1986)

 \bibitem{Bars} I. Bars, Phys.Lett. \underbar{B403}, 257, (1997), hep-th/9704054 
 \bibitem{Neeman}   Y. Neeman, Rev.  Mod. Phys.  \underline{37}, 227, (1965)
\bibitem{Joseph}  D. Joseph, Rev.  Mod. Phys.  \underline{37}, 225, (1965)
 
 

 \bibitem{RS}L. Randall \& R. Sundrum, 
Phys.Rev.Lett. \underbar{83}, 4690, (1999) hep-th/9906064 
\bibitem{Lupercio} E. Lupercio \&  B. Uribe,  math.DG/0402318
\bibitem{Kawamura} Y. Kawamura,  hep-ph/9902423
 \bibitem{Anne}Anne-Christine Davis et al, Phys. Lett \underbar{B504}, 254, (2001), hep-ph/0008132
  \bibitem{MaiaFr}  M.D. Maia  et al, Int. J. Mod. Phys. \underline{A17}, 4341,(2002)
 \bibitem{Carter} R. Battye \&  B. Carter,  Phys. lett. B, \underline{509}, 331, (2001)
\bibitem{Will}  C. M. Will,  {\em  Theory and  Experiment in Gravitational Physics}, revised  ed.  Cambridge,  (1993)
 \end{thebibliography}
\end{document}